\documentclass[a4paper,10pt,showpacs,pra,twocolumn]{revtex4}
\usepackage{graphicx}
\usepackage{amsmath}            % volle Mathematik
\usepackage{amstext}
\usepackage{amssymb}
\usepackage{subfigure}
\newcommand{\bra}[1]{\langle #1|}
\newcommand{\ket}[1]{|#1 \rangle}
\newcommand{\exponent}[1]{\ensuremath{\mathrm e^{#1}}}
\newcommand{\I}{\ensuremath{\mathrm{i}}}
 
\newcommand{\braket}[2] {\langle #1|#2 \rangle}

\newcommand{\partdiff}[1]{\frac{\partial }{\partial #1}}

\begin{document}
\title{The non-Abelian bosonic quantum ring}

\author{Michael Merkl$^1$, Gediminas Juzeli\=unas$^2$ and Patrik \"Ohberg$^1$}
\affiliation{$^1$\textit{SUPA}, Department of Physics, Heriot-Watt University, Edinburgh, EH14 4AS, United Kingdom}
\affiliation{$^2$Institute of Theoretical Physics and Astronomy of Vilnius University,
A. Gostauto 12, 01108 Vilnius, Lithuania }

\date{\today}

%%%%%%%%%%%%%%%
\begin{abstract}
We investigate the dynamics of a  spinor Bose-Einstein condensate which is governed by an optically induced non-Abelian gauge potential. 
Using a ring shaped trap to confine the atoms and a hydrodynamic ansatz, nonlinear Josephson type  equations are found to describe the system. The degenerate eigenstates which show rotation are solved exactly. We consider a homogenous filled ring and observe population dynamics between the two quasi-spin components but also space dependent Josephson oscillations. Stable mass currents can be observed which are induced by the constant non-Abelian effective magnetic field in the limit of weak interactions. For strong interactions the appearance of two-component dark soliton-like objects are observed.\end{abstract}
\maketitle

\section{Introduction}

The quantum ring is the Drosophila of quantum physics. The periodic boundary conditions have provided in many cases analytical solutions for non-trivial many-body problems and topological excitations such as persistent currents.
The atomic Bose-Einstein condensate (BEC) \cite{ReviewBECDalfovo,Dalibard2008r} has in this respect proven to be a remarkably versatile tool for studying a plethora of fundamental quantum phenomena. The Josephson effect is a prime example, where coherent dynamics takes place between two quantum states.
Josephson oscillations in ultra cold atomic systems where indeed soon after the
first realisations of a BEC \cite{Science.269.198,PhysRevLett.75.1687,PhysRevLett.75.3969} in the focus of interest. Characterising the 
new condensed state lead directly to the question how two condensates
interact. The Josephson effect of a weakly coupled macroscopic wave
function in superconductors gives a well understood archetype \cite{ReviewBECDalfovo, ReviewBECLegget}. For ultracold atoms there are two different scenarios available: a single component BEC which is separated by a small tunnelling barrier in a double well potential \cite{Oberthaler_2005},
 and a two component spinor gas trapped in a single trap with an internal coupling between the two condensates \cite{Williams_1999, Internal_Patrik}. 

The Josephson effect is a general one. It is a manifestation of phase coherence between two coupled quantum systems. Quantised vortices is another phenomenon which relies on many-body phase coherence, and  a paradigm for BECs \cite{ReviewBECDalfovo}. 
There are several techniques how to nucleate vortices in a BEC.
With the phase imprinting technique, which is widely used to create dark solitons \cite{PhysRevLett.83.5198}, a
phase with the wanted properties is optically induced into the gas \cite{PhysRevA.60.R3381,PhysRevLett.79.4728,kanamoto:063623}.
The most intuitive way, stirring with a laser beam \cite{Raman_2001,  Damski_2002} and rotating the trap \cite{Chiba_2008, Lundh_2003, Food_2001}, has also been used to create vortices in experiments. Stirring with a laser can be a harsh tool if a small number of particles are used, as is often the case if the quantum Hall regime is the goal. Dynamic instabilities are the main mechanism behind the vortex nucleation when reaching the critical rotation frequency \cite{Sinha_2001} above which the vortices appear. Rotating the trap leads to a homogeneous effective magnetic field in the equation of motion of the gas in the rotating frame. A different approach uses optically induced effective magnetic 
fields which is a non-rotating setup where vortices 
\cite{Murray_2007,PhysRevA.79.011604,Lin2009,SpilmanNature09,ZwierleinNature2009a} or solitons \cite{Juzeliunas2007L} can be created too. Without stirring or phase imprinting, vortex nucleation has also been observed in a mixing experiment of multiple trapped condensates \cite{scherer:110402}. 
To date most of the experiments have been done with single component condensates, but vortices are also
 a feature of spinor condensates \cite{woo:031604}.

In this paper we investigate a spinor condensate in a 1D ring which is subject to an optically induced non-Abelian magnetic field. Artificial gauge fields have attracted considerable interest lately, mainly due to the intriguing analogies which they provide between ultracold quantum gases and magnetic effects in solid state and condensed matter systems. The non-Abelian gauge potentials can be created by laser assisted tunnelling \cite{ruostekoski2002,osterloh2005} in optical lattices, or by relying on optically induced degenerate dark state dynamics \cite{Ruseckas_2005d}. 

We are interested in the dynamics of a quasi-spin in the ring. The presence of a matrix gauge potential which couples to the quasi-spin provides the machinery for exotic topological states. In particular we show that the simplest non-trivial gauge potential which is also non-Abelian, and which has been used extensively in recent papers \cite{Stanescu2007,Juzeliunas_2007a,Jacob_2007,Merkl_2008a,vaishnavZB,Merkl_2009a}, can give rise to dynamically induced local spin precession.

The paper is organised as follows.
We start by introducing the equations of motions which govern the dynamics of the atomic quasi-spin, followed by an exact solution of the eigenvalue problem for the system.
In the following section we use the eigenstates and a hydrodynamic ansatz for the BEC to derive a set of equations which resemble those who describe 
the internal Josephson effect.  In Section V we investigate the spacial density oscillations and discuss the resulting atomic spin currents in the ring. Finally we conclude with a discussion about the influence of collisional interactions on the dynamics. 

%%%%%%%%%%%%%%%%%%%%%%%%%%%%%%%%%%%%%%%%%%%%%%%%%%%%%%%%%%%%%%%%%%%%%%%%%%%%%%%%%%%%%%%%%%%%%%

\section{The ring geometry and the origin of the gauge potentials}

\begin{figure} [t,h,b]
	\includegraphics[width=0.45\textwidth]{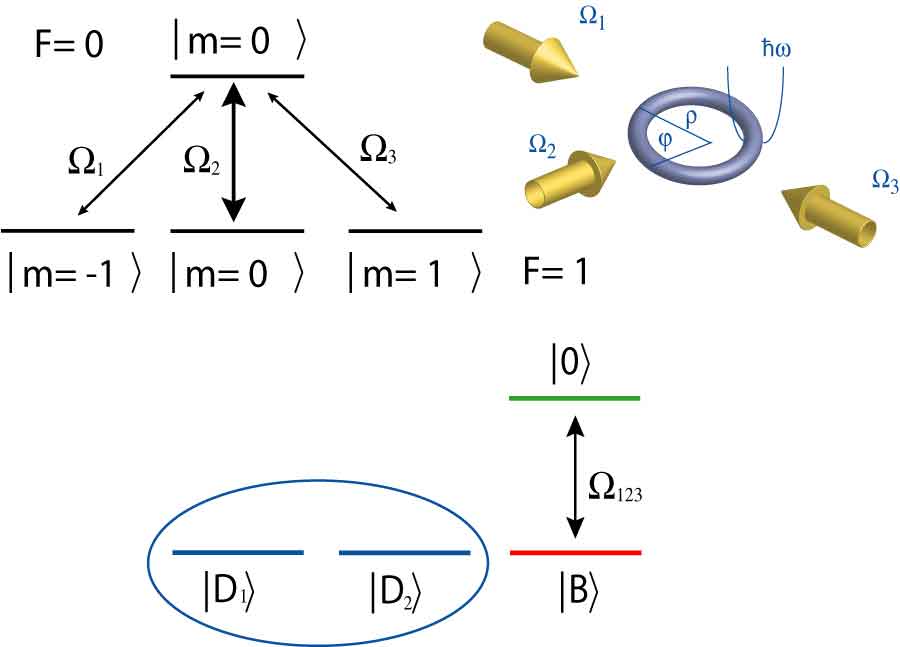}
	\caption{ \label{pic-exp-setup} (Upper left) atomic bare tripod level structure and (upper right) geometrical setup with three resonant laser beams to generate a non-Abelian gauge field in the dressed state basis (below). The dark state sub-manifold is encircled.
 The lasers drive the system with  Rabi frequencies $\Omega_1 =  \sin \theta_0 \, \Omega\exponent{- \I \kappa x} /\sqrt{2}$,
	$\Omega_2 =  \sin \theta_0 \,\Omega \exponent{ \I \kappa x} /\sqrt{2}$, $\Omega_3 = \cos \theta_0 \Omega \exponent{ \I \kappa y}$, where $\cos \theta_0 = \sqrt{2}-1$ and 
	$\Omega = \sqrt{  \sum_{\rm{i=1}}^3  | \Omega_{\rm{i}} |^2 } $. (See \cite{Juzeliunas_2007a} for more details.)
	} 
 \end{figure}

In the following we consider a Bose gas of atoms which are described by a tripod level scheme \cite{Ruseckas_2005d,PhysRevA.59.2910} (Fig. 1). 
The tripod scheme can for instance be formed by three ground $F=1$ states ($m_F=0,\pm 1$) and an excited $F=m_F=0$ state. 
In particular the transition $5S_{1/2} (F=1) \leftrightarrow 5P_{3/2} (F=0)$ in $^{87}$Rb or the transition
$2^3S_1 \leftrightarrow 2^3P_0$ in ${^4}$He$^*$ could be used. Three lasers are connecting the ground-state levels $m_F=-1$, $0$ and $1$ to the excited state 
with  $\sigma_+$, $\pi$, 
and $\sigma_-$ polarisations, with Rabi frequencies $\Omega_{1,2,3}(\mathbf{r})$, and phases $S_{1,2,3}(\mathbf{r})$.
The resulting atom-light interaction leads to two eigenstates of zero energy, called dark states $|D_{1,2}(\mathbf{r})\rangle$, which are linear superpositions of the three lowest bare states.
If ${\Omega}= \sqrt{ \sum_i |\Omega_i|^2 }$ is large compared to any other energy scale, including laser detuning, Doppler and Zeeman shifts and interaction energy,  
we can safely neglect transitions out of the dark state manifold. An atom prepared in such a dark state can be described by the general state $|{\Psi(\mathbf{r},t)}\rangle = \sum_{i=1}^2 \Psi_i(\mathbf{r},t) |{D_i(\mathbf{r})}\rangle$, where 
$\Psi_i (\mathbf{r},t)$ is the wave function for atoms in $|D_i\rangle$.  
If we allow the atom to move, {\it i.e.}, we add a kinetic energy to the light-matter interaction Hamiltonian,  and project onto the dark state manifold, we obtain the effective Schr\"odinger equation~\cite{Ruseckas_2005d}
\begin{equation} \label{eq schroedinger equation effectiv1}
 i \hbar \frac{\partial}{\partial t} {\Psi} = \Big[ \frac{1}{2m} \left(\mathbf{p}-\mathbf{A}\right)^2 + \mathbf{{V}}+ 
\mathbf{{\Phi}} \Big] {\Psi},
\end{equation}
where ${\Psi}^T =(\Psi_1,\Psi_2)$, $\mathbf{p}$ is the momentum operator and $m$ the atomic mass. The key point here is that the spatial dependence of the laser arrangement leads to an effective matrix vector potential \cite{Berry_1984,Wilczek-Zee_1984,Mead_1992a}
${\bf A}_{nm}=i\hbar\langle D_n({\bf r})|\nabla D_m({\bf r}) \rangle$. In addition, ${\bf V}_{nm}= \bra{D_n} \sum_{i=1}^{3} V_i (\mathbf{r}) \ket{D_m}$ and ${\bf \Phi}_{nm}=(\hbar^2/2m)\langle D_n({\bf r})|\nabla B({\bf r}) \rangle\langle B({\bf r})|\nabla D_m({\bf r}) \rangle$ 
are effective scalar potential matrices, where $|B({\bf r})\rangle$ is the so-called bright state, {\it i.e.}, the linear combination of the ground levels which 
does couple with the excited state. The potential ${V_{i}}$ 
may include any detuning of the n-th laser from the resonant transition. 
The gauge potential can be shaped in quite arbitrary manner, but already a very simple configuration
as depicted in fig. \ref{pic-exp-setup} leads to a non-trivial, non-Abelian potential \cite{Juzeliunas_2007a,Jacob_2007}.

%%%%%%%%%%%%%%%%%%%%%%%%%%%%%%%%%%%%%
%%%%%%%%%%%%%%%%%%%%%%%%%%%%%%%%%

% INTERACTION + GROSS-PITAEVSKII TREATMENT

% BEC

The term $\mathbf{p}\cdot{\mathbf A}$ describes spin-orbit coupling. It will give rise to two single atom dispersion branches. For a homogeneous 2D system degenerate minima in the dispersion 
may occur, which would for the many-body system cause fragmentation and preclude the formation of a BEC \cite{Stanescu2007}. This is not necessarily the case in a 1D ring, as we will show later, even if the ring of course resides in the two dimensional plane.  

% Interactions

Inter-atomic interactions play an important role in the properties of the Bose gas at low temperatures. We will consider the case where the interaction energy is much smaller than the Rabi frequency $\Omega$, which guarantees that we always stay in the dark-state manifold. It is important to note that the 
ground states $|j=-1,0,1 \rangle$ form a spin-1 Bose gas and that short-range interactions are dominantly $s$-wave. Because of  symmetry collisions occur only in two 
different channels with a total spin of $0$ and $2$. These collisions are characterised by the corresponding $s$-wave scattering  lengths $a_0$ and $a_2$~\cite{Ho1998}. 
These scattering lengths are in principle different, although in practice they are very similar. For simplicity of the discussion we consider $a_0=a_2=a$. This allows us to prevent spin flip collisions and we
%%%%%%%%%%%%%%%%%%%%%%%%%%%%%%%%%%
us a Gross-Pitaevskii mean field description. The dark state spinor gas is described by a two component wave function
	\begin{equation} \label{eq-def-spinor1}
	{\Psi}= 
	\left( \begin{array}{c} 
	\Psi_1 \\
	\Psi_2 
	\end{array} \right)	
	\end{equation}
which gives a Hamiltonian of the form
	\begin{equation} \label{eq-hamiltonien1}
	H= \frac{1}{2m} \big( \mathbf{p}-\mathsf{A} \big)^2 + V(\mathbf{r}) + g' (|\Psi_1|^2+|\Psi_2|^2).
	\end{equation}
Here, $g'=4 \pi \hbar^2 a N/m $, where $a$ is the s-wave scattering length discussed above, and $N$ is the particle number with $\int \; \text{d} V (|\Psi_1|^2+|\Psi_2|^2)=1$.  
The vector potential $\mathsf{A}$ can be chosen quite arbitrary.  We are 
here interested in a constant, non-Abelian potential in terms of Pauli matrices $\sigma$ \cite{Juzeliunas_2007a,Stanescu2007}
	\begin{equation} \label{eq-def-A-cartesian}
	\mathsf{A} = - \hbar \kappa (\sigma_x \mathbf{e}_x + \sigma_y \mathbf{e}_y).
	\end{equation}
The Pauli matrices reflect the spacial dependence while $\kappa$ 
is determined by the wave-vectors  of the underlying laser fields.
The Schr\"odinger equation 
for the the centre of mass  spinor  motion then becomes with $I=g' (|\Psi_1|^2+|\Psi_2|^2)$
	\begin{eqnarray} \label{eq-def-schroedinger-ansatz}
	&&\I \hbar \partdiff{t} 
	\left( \begin{array}{c} 
	\Psi_1 \\
	\Psi_2 
	\end{array} \right) =  \\	
	&&
	\left( \begin{array}{cc} 
	-\frac{ \hbar^2 \nabla^2_{\mathbf{r}} }{ 2 m } + \frac{\hbar^2 \kappa^2}{m} +I
	&    \frac{\hbar^2 \kappa  }{2 m } (-\I \partial_x - \partial_y) \\
	 \frac{\hbar^2 \kappa  }{  m } (-\I \partial_x +  \partial_y) 
	&
	-\frac{ \hbar^2 \nabla^2_{\mathbf{r}} }{ 2 m } + \frac{\hbar^2 \kappa^2}{m} +I
	\end{array} \right) 
	\left( \begin{array}{c} 
	\Psi_1 \\
	\Psi_2 
	\end{array} \right).	\nonumber
	\end{eqnarray}

In this paper we will restrict ourselves to a 1D situation on a ring.  Experimentally an effective 1D ring
for cold atoms can be achieved by strong harmonic trapping in the transversal direction \cite{Hofferberth_2007}.
In 2D it has been shown 
 that the non-Abelian dynamics is all but trivial \cite{Jacob_2007,Stanescu2007}. In this work, however, we investigate a ring geometry where the centre of mass dynamics is 1D but the corresponding spin dynamics is not, see Figure \ref{pic-exp-setup}. There are a number of techniques available to experimentally create sophisticated ring traps  \cite{arnold:041606,GRIFFIN_2008},
where even toroidal-shaped traps for BECs have been proposed \cite{fernholz:063406}. 

We describe the ring dynamics using polar coordinates with the azimuthal angle $\varphi$ and  the radius $\rho$. The gauge potential in Eq.  (\ref{eq-def-schroedinger-ansatz}) is consequently transformed as 
	\begin{eqnarray}
	&&\mathsf{A}_{\rho}= -\hbar \kappa ( \sigma_x \cos \varphi + \sigma_y \sin \varphi ) \\
	&&\mathsf{A}_{\varphi}=- \hbar \kappa (- \sigma_x \sin \varphi + \sigma_y \cos \varphi) \\
	&&\mathsf{A}_{z}=0.
	\end{eqnarray}
The radial and $\mathrm{z}$ components are neglected in the following. The ring trap 
provides with $ \hbar \omega_{\rho} \gg \frac{1}{2m} (\mathbf{p}-\mathsf{A})_{\rho}^2 $ the dominating energy scale and hence any dynamics in these directions are suppressed by the harmonic oscillator ground state with a cross section $\bar{s}= \hbar / m \omega_\rho$. 

The relevant vector potential is now of the form
	\begin{equation} \label{eq-def-A-ring2}
	\mathsf{A} _\varphi= - \hbar \kappa
	\left( \begin{array}{cc} 
	0 & -\I \exponent{-\I \varphi}\\
	\I \exponent{\I \varphi} & 0 
	\end{array} \right) \mathbf{e}_\varphi,	
	\end{equation}
while the momentum operator in $\mathsf{e}_{\varphi}$-direction reads
	\begin{equation}
	\mathbf{p}_\varphi = -\I\hbar \frac{1}{\rho} 
	\partdiff{\varphi} \mathbf{e}_\varphi. 
	\end{equation}

The gauge potential now depends of the angle $\varphi$.
This $\varphi$-dependence leads to additional factors compared to the expression for the gauge potential in Eq. \eqref{eq-def-schroedinger-ansatz}. 
The resulting  Hamiltonian  becomes 
	\begin{equation} \label{eq-angular-hamilton-matrix1}
	H_\varphi 
	 = 
	\left( \begin{array}{cc} 
	-\frac{  \partial^2_{\varphi} }{ 2} + \kappa^2 + V + g |\Psi|^2
	&\exponent{-\I \varphi}  \Big[ -\I \frac{  \kappa }{ 2  } + { \kappa \partial_{\varphi} }{ }\Big] \\
	\exponent{\I \varphi} \Big[ -\I \frac{  \kappa }{ 2  } - { \kappa \partial_{\varphi} }{  }\Big] & -\frac{ \partial^2_{\varphi} }{ 2  } + { \kappa^2}  - V + g |\Psi|^2
	\end{array} \right),	
	\end{equation}
expressed in units of $\frac{ \hbar^2 }{ m \rho^2}$ for the energy, length in units of $\rho$, time in units of $\frac{m \rho^2}{\hbar}$, and $g= \frac{4 \pi a N }{s_\perp}$, where $N$ is the number of particles and $s_\perp = \frac{ \hbar }{m \omega_\perp} \frac{1}{\rho^2}$ the transverse cross section.
The first term in the off-diagonal coupling vanishes in the limit $\rho \to \infty$ and 
can be seen as a kind of centrifugal term due to circular motion.  The potential $V$ is a constant external offset in the trapping potential of the two different dark states which can be controlled by detuning the incident laser beams from the resonant transitions in the tripod level structure.

\section{Eigenstates} \label{sec-Eigenstates}

%%%%%%%%%%%%%%%%%%%%%%%%%%%%%%%%%%%%%%%%%%%%%%%%%%%%%%%%%%%%%%%%%%%%%%%%%%%%%%%%%%%
\begin{figure} [b,t,h]
	\includegraphics[width=0.45\textwidth]{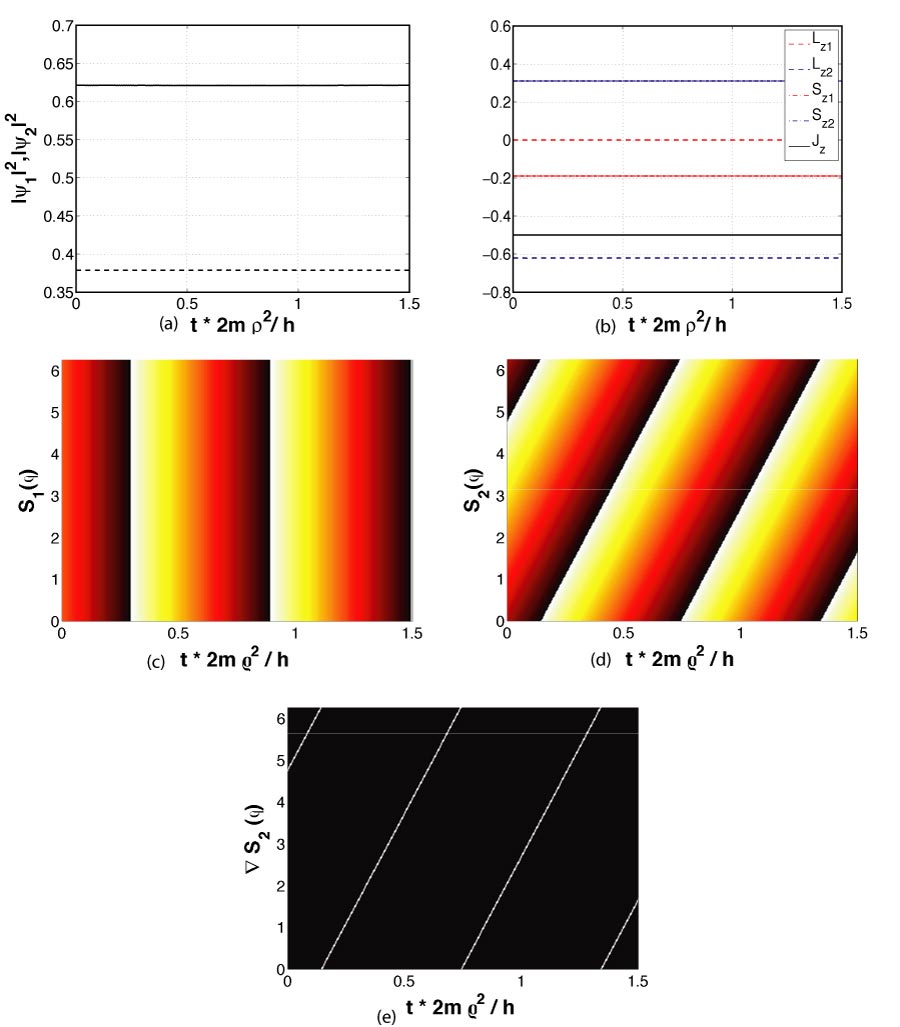}
 	\caption{ \label{pic-eigen-ring} The eigenstate $\Psi^+_{q=0}$: (a) Population, (b) total ($J_z$), orbital ($L_z$) and spin angular momentum ($S_z$), (c,d) the phase of the spin components and (e) the gradient of the phase  as a function of time (here for  $q=0$, $\rho= 2 \kappa^{-1}, \kappa=1$). The white line in (e) shows the $2\pi$-phase jump and indicates the quantised winding number of the second spinor component. 
	} 
 \end{figure} 
%%%%%%%%%%%%%%%%%%%%%%%%%%%%%%%%%%%%%%%%%%%%%%%%%%%%%%%%%%%%%%%%%%%%%%%%%%%%%%%%%%%

The eigenstates of the non-Abelian ring can be calculated analytically using
the Heisenberg equation of motion for the velocity operator. The interaction term is merely a function of $\mathbf{r}$ and gives no contribution to 
	\begin{eqnarray} \label{eq-def-velocity}
	\mathbf{\hat{v}} &=& -\I \big[ \hat{H},\mathbf{\hat{r}} \big] 
	=(\mathbf{p}-\mathsf{A}).
	\end{eqnarray}
For a homogeneously filled ring where $[\mathbf{p},|\Psi|^2]=0$,  the Hamiltonian (\ref{eq-angular-hamilton-matrix1}) commutes with $\mathbf{\hat{v}}$. The common eigenstates are found from the eigenvalue equation for the velocity operator which has the form
	\begin{equation} \label{eq-chring-eigenvelocity}
	(-i   \partdiff{\varphi} - \mathsf{A} ) \Psi =  K \Psi,
	\end{equation}
 with the eigenvelocity $ K$. The Hamiltonian satisfies the eigenvalue equation
 	\begin{equation}
	H_\varphi \Psi = E_K \Psi
	\end{equation}
 where the eigenenergy is given by $E_K = K^2$. 
 To solve the eigenvalue problem (\ref{eq-chring-eigenvelocity}) we use the ansatz 
	\begin{equation} \label{eq-ansatz-eigenstate}
	\Psi_q(\varphi)=
	\left( \begin{array}{c} 
	\exponent{\I q \varphi} \tilde{\Psi}_{q1}\\
	\exponent{\I (q+1) \varphi} \tilde{\Psi}_{q2}  
	\end{array} \right),
	\end{equation}
where $q$ is an integer.
Eq. (\ref{eq-ansatz-eigenstate}) leads to a set of two equations for the remaining components $\tilde{\Psi}_{qi}$,
	\begin{eqnarray}
	K  \tilde{\Psi}_{q1} &=& q \tilde{\Psi}_{q1 }+\I \kappa  \tilde{\Psi}_{q2}  \label{eq-estat-aux1},\\
	K  \tilde{\Psi}_{q2} &=& - \I \kappa \tilde{\Psi}_{q1}+ (q+1) \tilde{\Psi}_{q2} \label{eq-estat-aux2}  ,
	\end{eqnarray}
with a non-trivial solution if
	\begin{equation}
	\left| \begin{array}{cc} 
	K  - q & -\I \kappa  \\
	\I \kappa  & K - q - 1  
	\end{array} \right| =0.
	\end{equation}
The determinant gives the condition
	\begin{equation} \label{eq-cond-eigen1}
	(K  - q)(K -q-1)-\kappa^2 =0.
	\end{equation}
For convenience we introduce the dimensionless quantity $u$ as
	\begin{equation} \label{eq-def-u1}
	\kappa^2  =(u+1/2)^2-1/4=u(u+1).
	\end{equation}
After inserting Eq. (\ref{eq-def-u1}) in (\ref{eq-cond-eigen1}) we get for
the eigenvelocities 
	\begin{eqnarray} \label{eq-evelocities1}
	K_{+} =q + u+ 1, \quad K_{-}  = q-u.
	\end{eqnarray}
 From Eqs. 
(\ref{eq-estat-aux1}) and (\ref{eq-evelocities1}) we get
	\begin{equation} \label{eq-ratio-eigenstates}
	\I \kappa  \tilde{\Psi}
^{\pm}_{q2} = \Big( \frac{1}{2} \pm (u+\frac{1}{2}) \Big) \tilde{\Psi}^{\pm}_{q1}.
	\end{equation}
The eigenstates are now
	\begin{equation} \label{eq-def-eigenstate-ring}
	\Psi_q^{\pm} = 
	\left( \begin{array}{c} 
	c \sqrt{u + 1/2 \mp 1/2} \, \exponent{\I q \varphi} \\
	\mp \I c\sqrt{u+ 1/2 \pm 1/2} \, \exponent{\I (q+1) \varphi}   
	\end{array} \right),
	\end{equation}
where $c=(u+1/2)^{-1}$ is a normalisation constant. From these expressions we see that each spinor component has a winding number associated to it, and that the winding number is always shifted by one between the two eigenstate components.
The ratio $\tilde{\Psi}_{q1}/\tilde{\Psi}_{q2}$ does not depend on the quantum number $q$ nor the interaction strength $g$ but on the vector potential $\kappa$.
For any quantum number $q$ the  gradient of the phase difference is $|\nabla S_2 - \nabla S_1 |= \varphi$.

\begin{figure}[h,t,b]
	\includegraphics[width=0.45\textwidth]{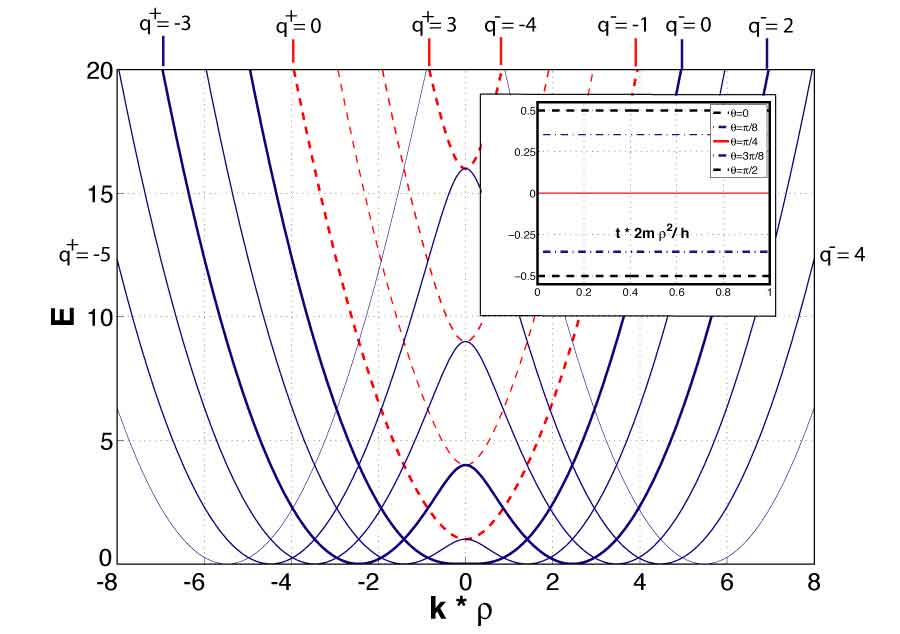}
	\caption{ \label{pic-energy} Energy spectrum of the ring. The dashed (red) curves belong to $q^+ \ge 0$ and are degenerate with $q^- < 0$. The solid (blue) curves belong to $q^- \ge 0$ and $q^+ <0$. The $q^{\pm}$ is the states winding number and the $\pm$ indicates the branches of $E^{\pm}$. All energies are two times degenerate. Superpositions $\sin \theta \ket{\Psi^-_{q_0}} + \cos \theta \ket{\Psi^+_{-q_0-1}}$ have different $J_z$ as shown in the insert for $q_0=0$.}
\end{figure}

The eigenenergies $E^{\pm}$ of the eigenstates $\Psi^{\pm}$ are calculated using (\ref{eq-evelocities1}) together with (\ref{eq-def-u1}),

	\begin{equation}
	E^{\pm}  = K^2 = \Big[ q + 1/2 \pm \sqrt{ \kappa^2  + 1/4} \Big]^2.
	\end{equation}
There are two branches $E^{\pm}$ for every value of $q$ which can be positive or negative. The states with $E^{\pm}$ and $q=q_0$ are degenerate with $E^{\mp}$ and $q=-q_0-1$ 
as  shown in Fig. \ref{pic-energy}. $\ket{\Psi_{q_0}^-}$ and $\ket{ \Psi_{-q_0-1}^+}$ span a space of infinite many groundstates. 
However, the orthogonal eigenstates $\ket {\Psi^{\pm}_q }$ are eigenstates of the total angular momentum operator $J_\mathrm{z}= L_\mathrm{z} + S_\mathrm{z}= -\I \partial_\varphi + \frac{1}{2} \sigma_\mathrm{z}$
as they fulfil   $J_\mathrm{z} \ket{  \Psi^{\pm}_q }= \pm (q+\frac{1}{2}) \ket{ \Psi^{\pm}_q }$. 
Hence all superpositions of the form $\alpha \ket{\Psi_{q_0}^-}+ \beta \exponent{\I \xi} \ket{\Psi_{-q_0-1}^+} $ differ in their total angular momentum $(\alpha^2 - \beta^2)  (q_0 + 1/2)$, where the real numbers $\alpha, \beta$ obey $\alpha^2+\beta^2=1$ and $\exponent{\I \xi}$ reflects just a constant phase factor between the two states.

For the ring  $J_\mathrm{z}$ is a conserved quantity  as  $J_\mathrm{z}$ and $H$ have common eigenstates. 
If the condensate is first prepared in the ring and then subsequently the lasers are switched on, the conservation of the total angular momentum should prevent the system from fragmentation \cite{mueller:033612}.

%%%%%%%%%%%%%%%%%%%%%%%%%%%%%%%

%%%%%%%%%%%%%%%%%%%%%%%%%%%%%%%%%%%%%%%%%%%%%%%%%%%%%%%%%%%%%%%%%%%%%%%%%%%%%%%
\section{The Internal Josephson Effect}

%%%%%%%%%%%%%%%%%%%%%%%%%%%%%%%%%%%%%%%%%%%%%%%%%%%%%%%%%%%%%%%%%%%%%%%
\begin{figure} [t,h,b]
 	\includegraphics[width=0.45\textwidth]{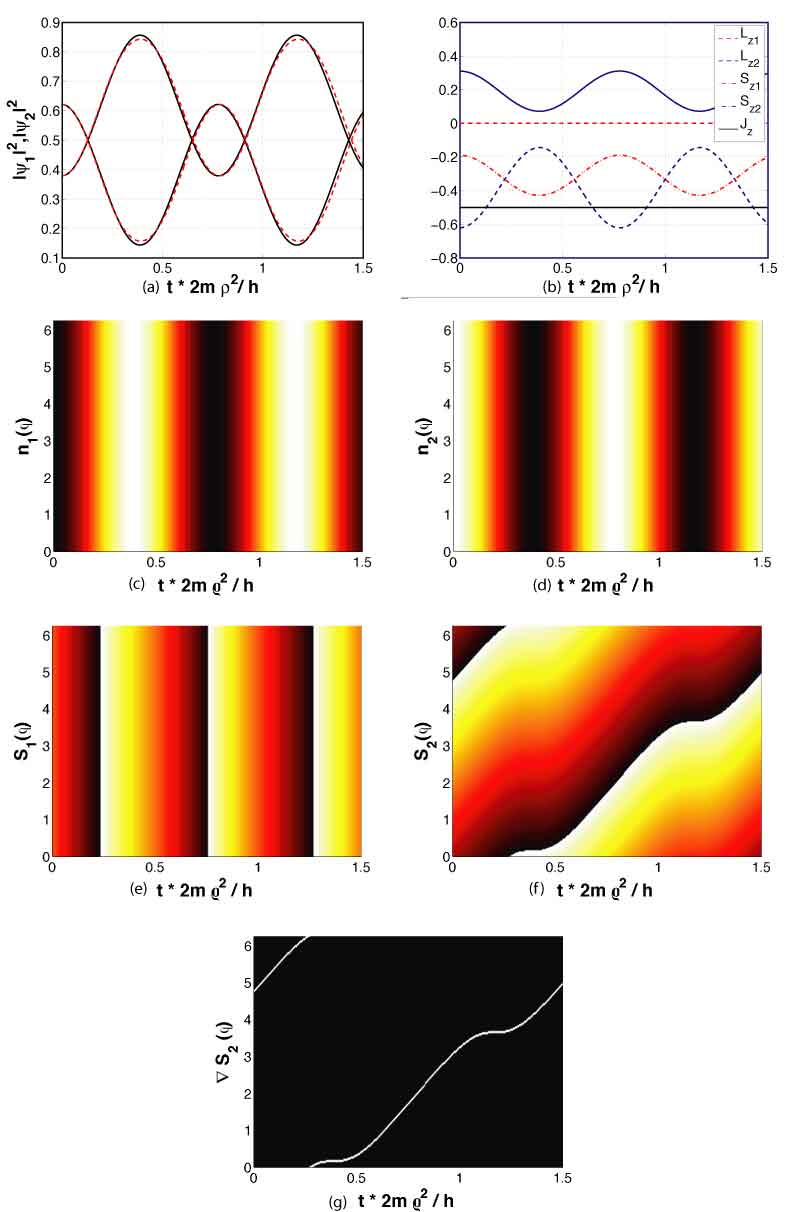}
 	\caption{ \label{pic-J-AC} The Internal Josephson effect by applying an external potential $V=0.5$: (a) The population of the two quasi-spin components is shown. The black and solid curve shows the population simulated numerically while the red and dashed line is the analytic approach from Eqs. \eqref{eq-Josefson-25}. (b) shows total ($J_z$), orbital ($L_z$) and spin angular momentum ($S_z$). In  (c,d) we plotted the density of the BEC, the minima are black and the density maxima are white. (e,f) show the phase and (g) the phase gradient $\nabla S_2$ of the second component for the AC-Josephson effect. 
In all cases $\rho = 2 \kappa^{-1}$ with an initial eigenstate $( c\sqrt{u} \exp(0), -\I c\sqrt{u+1} \exp(\I \varphi) \; )$ }\end{figure} 
%%%%%%%%%%%%%%%%%%%%%%%%%%%%%%%%%%%%%%%%%%%%%%%%%%%%%%%%%%%%%%%%%%%%%%%%%%%%%%%%%%%%%%%%%%%%%%%%%%%%%% 
In the following we will use a hydrodynamic description of the condensate \cite{PhysRevLett.77.2360} to obtain a set of equations describing the dynamics on the ring. We write an ansatz for the wavefunction as
	\begin{equation} \label{eq-ansatz-josephson}
	\Psi_J = 
	\left( \begin{array}{c} 
	\Psi_{J1}\\
	\Psi_{J2}  
	\end{array} \right)
	=
	\left( \begin{array}{c} 
	\sqrt{n_1} \exponent{\I S_1 }\\
	-\I \sqrt{n_2} \exponent{\I S_2 }  
	\end{array} \right),
	\end{equation}
with the (normalised) local density $n_i (\varphi)$ and the phase $S_i(\varphi)$.  

By using the transformation in Eq. \eqref{eq-ansatz-josephson} together with the Hamiltonian from  \eqref{eq-angular-hamilton-matrix1}, the time dependent Schr\"odinger equation is rewritten
by multiplying with the hermitian conjugate $\Psi_J^\dagger$ from the left, 
	\begin{equation} \label{eq-ansatz-hydro}
	\Psi_J^\dagger \I  \partial_t \Psi_J = \Psi_J^\dagger H_\varphi \Psi_J.
	\end{equation}
From Eq. \eqref{eq-ansatz-hydro} we get with a straightforward calculation (see appendix \ref{Appendix-Jos})  after separating in real and imaginary parts 
the coupled nonlinear equations of motion for the phases $S_i (\varphi, t)$ and the local densities $n_i (\varphi, t)$:
	\begin{eqnarray}
	 \dot{S}_1 &=&  [ \frac{(\partial_\varphi n_1)^2}{8  n_1^2 }  + \frac{\partial_\varphi^2 n_1}{2  n_1} - \frac{1}{2  }(\partial_\varphi S_1)^2  \nonumber  \\
	&&   - (\kappa^2+V + g(n_1+n_2 ) ]  \nonumber\\
	&& - \frac{ \kappa\sqrt{n_2} }{ \sqrt{n_1} } \Big( (  \partial_\varphi S_2   -\frac{1 }{2 }  )\cos \Delta  +   \frac{ \partial_\varphi n_2  }{ 2 n_2 }  \sin  \Delta \Big)   \label{eq-Josephson-long1}  \\
	\dot{n}_1 &=&  \Big( - \partial_\varphi S_1 \partial_\varphi n_1 - n_1 (\partial_\varphi^2 S_1)  \Big) +  2 \kappa\sqrt{n_1n_2}
	\label{eq-hydro-1-density}  \nonumber  \\
	&&   \times  \Big(   (   \partial_\varphi S_2 - \frac{1 }{2 } ) \sin \Delta -  \frac{\partial_\varphi n_2   }{ 2 n_2 }  \cos \Delta\Big)  
\end{eqnarray}

\begin{eqnarray}
	\dot{S}_2 &=&  [ \frac{(\partial_\varphi n_2)^2}{8  n_2^2 }  + \frac{\partial_\varphi^2 n_2 }{2  n_2} - \frac{ (\partial_\varphi S_2)^2}{2 } \nonumber \\
	&&   - (\kappa^2-V + g (n_1+n_2 )) ]  \nonumber\\
	&& - \frac{ \kappa\sqrt{n_1} }{ \sqrt{n_2} } \Big(  (  \frac{1 }{2 } +  \partial_\varphi S_1 )\cos \Delta -   \frac{ \partial_\varphi n_1  }{ 2 n_1 }  \sin  \Delta\Big)  \label{eq-Josephson-long3} 
	\end{eqnarray}
	
	\begin{eqnarray}
	\dot{n}_2 &=&  \Big( - \partial_\varphi S_2 \partial_\varphi n_2 - n_2 (\partial_\varphi^2 S_2)  \Big) +2 \kappa \sqrt{n_1n_2}
	\label{eq-hydro-2-density}\nonumber   \\
	&&    \times  \Big(  (    \partial_\varphi S_1-  \frac{1 }{2 }  ) \sin \Delta -  \frac{ \partial_\varphi n_1  }{ 2 n_1 }  \cos \Delta\Big)    , 
	\end{eqnarray}
where we have introduced $\Delta= S_2-\varphi-S_1$. For a completely filled ring with a constant density, $\partial_\varphi n_i =0$, the 
equations simplify considerably,
	\begin{eqnarray}
	  \dot{S}_1 &=&   -\frac{ 1}{2}(\partial_\varphi S_1)^2  - (\kappa^2 +V+ g (n_1+n_2 )) \nonumber \\
	 &&-  \frac{ \kappa\sqrt{n_2} }{ \sqrt{n_1} } (    \partial_\varphi S_2 -\frac{1 }{2 } )  \cos \Delta   \label{eq-Josephson-short1}\\
	 \dot{n}_1 &=&    - n_1 (\partial_\varphi^2 S_1)   
	  +    2 \kappa\sqrt{n_1n_2}( - \frac{1 }{2 } +  \partial_\varphi S_2 )  \sin \Delta     \label{eq-Josephson-short2}\\
	  \dot{S}_2&=&  [ - \frac{1}{2 }(\partial_\varphi S_2)^2   - (\kappa^2-V + g (n_1+n_2 )) ]  \nonumber \\
	 &&- \frac{ \kappa\sqrt{n_1} }{ \sqrt{n_2} }  (  \frac{1 }{2 } +  \partial_\varphi S_1 )\cos \Delta  \label{eq-Josephson-short3} \\
	 \dot{n}_2 &=&    - n_2 (\partial_\varphi^2 S_2)   +
	    2\kappa\sqrt{n_1n_2}  ( -  \frac{1 }{2 } +  \partial_\varphi S_1 )   \sin \Delta  \label{eq-Josephson-short4}
	\end{eqnarray}
The eigenstates \eqref{eq-def-eigenstate-ring} can be seen to fulfill  the relations $S_2 - \varphi - S_1=0$ and $\partial_\varphi^2 S_i=0$, hence $\cos \Delta =1$ and $\sin \Delta = 0$. The time evolution of the densities is then given by
\begin{eqnarray}
	&& \dot{n}_1 =   -  2 \kappa\sqrt{n_1n_2}  (  \frac{1 }{2 } -  \partial_\varphi S_2 )   \sin \Delta  =0 ,\nonumber \\
	&& \dot{n}_2 =   -   2\kappa\sqrt{n_1n_2} (   \frac{1 }{2 } -  \partial_\varphi S_1 )     \sin \Delta =0.\label{eq-population-time-ES}
	\end{eqnarray}
The phase difference $\Delta = S_2 - \varphi -S_1$, on the other hand, evolves according to
	\begin{eqnarray} \label{eq-Josefson-20}
	&&\dot{S}_1 - \dot{S}_2 =   \frac{1}{2 } \big(   (\partial_\varphi S_2)^2  -(\partial_\varphi S_1)^2\big) -2 V \label{eq-phase-difference-ES} \\
	&&- \kappa  \Big(   \sqrt{ \frac{ n_2} { n_1} } ( -\frac{1 }{2 } +  \partial_\varphi S_2 )    +\sqrt{ \frac{ n_1}  {n_2} }(  \frac{1 }{2 } +  \partial_\varphi S_1 )  \Big)  \cos \Delta \nonumber \\
	&&=   (\frac{1}{2} +  \partial_\varphi S_1) - (  \frac{1}{2}+\partial_\varphi S_1 ) =0, \label{eq-Josefson-21}
	\end{eqnarray}
where we have used the eigenstate properties $\partial_\varphi S_2 = 1 + \partial_\varphi S_1$ and $  \sqrt{ \frac{ n_1} { n_2}} - \sqrt{ \frac{ n_2} { n_1} }= \frac{1}{\kappa} $ and $V=0$.
%

%%%%%%%%%%%%%%%%%%%%%%%%%%%%%%%%%%%%%%%%%%%%%%%%%%%%%%%%%%%%%%%%%%%%%%%%%%%%%%%%%%%
\begin{figure} [b,t,h]
	\includegraphics[width=0.45\textwidth]{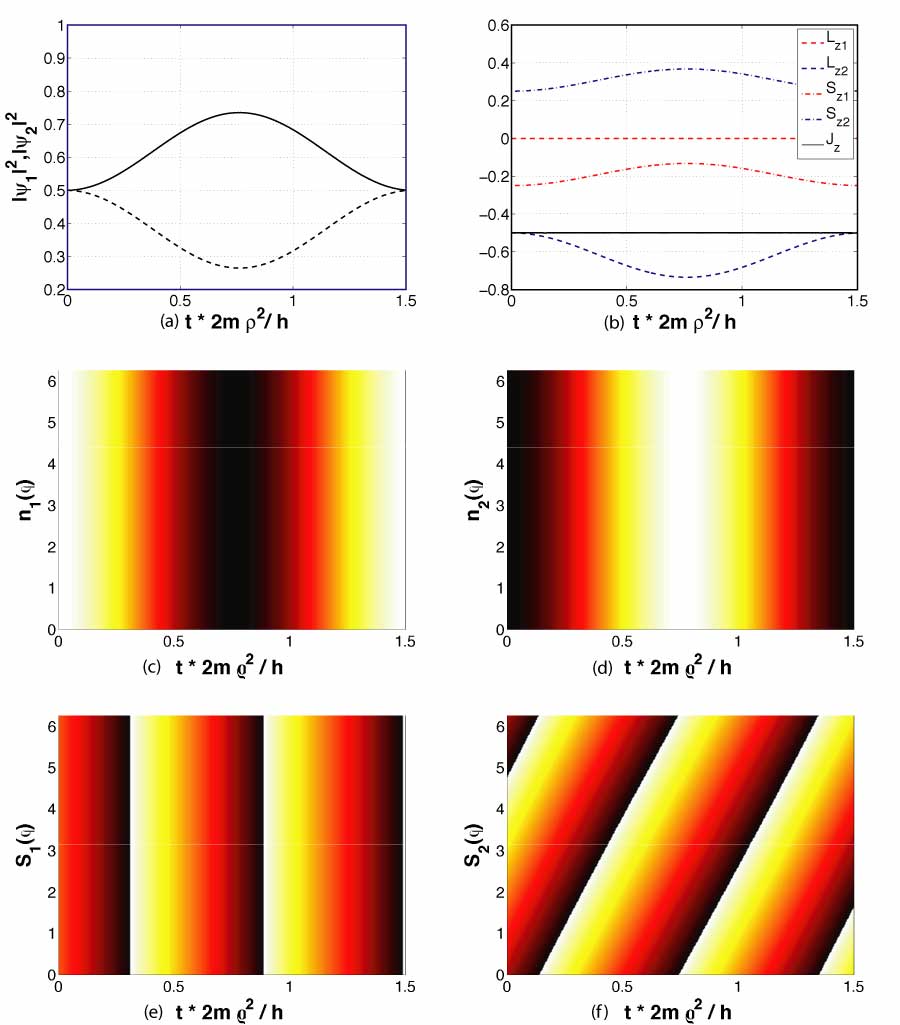}
 	\caption{ \label{pic-density-oscillations-ring} Internal Josephson effect for non eigenstate $(0.5 , -0.5 \I \exp(\I \varphi) \; )^{\rm T}$: (a) The quasi-spin population $\int_0^{2 \pi} \mathrm{d}\varphi \, |\Psi_i|^2$, $i=1,2$; (b) the total ($J_z$), orbital $(L_z)$ and spin angular momentum $(S_z)$ are plotted for each quasi-spin component. (c) and (d) show the density of component 1 and 2, where the minima are black and the maxima in white. (e) and (f) is the the phase of component 1 and 2, ranging over $\pm \pi$. 
	} \end{figure} 

Oscillations which are driven by the phase difference between two quantum fluids where first investigated in solid state physics using superconductors with a small tunnelling barrier, e.g., a grain boundary. The prediction of AC and DC currents in so called  superconducting interference devices (SQUIDS \cite{SQUID_Handbook}) was a great success of the BCS theory of cooper pairs. SQUIDS are nowadays widely used  e.g. in magnetic field sensors. 
The set of equations (\ref{eq-population-time-ES}) and (\ref{eq-phase-difference-ES}) describing the non-Abelian ring resembles the Josephson equation and does indeed give rise to coherent effects like population oscillations in the ring. 

If we apply a constant potential $\pm V$ to the two dark states, or if we change the population ratio $\tilde{\Psi}_{q1}/\tilde{\Psi}_{q2}$ of an eigenstate \eqref{eq-ratio-eigenstates}, the system shows population oscillations as illustrated in Figs. \ref{pic-J-AC} and  \ref{pic-density-oscillations-ring} for a non-interacting system with $g=0$.
The total angular momentum $J_\mathrm{z}$ is always conserved, whereas the oscillations of $L_{\mathrm{z}i}(t) = \bra{\Psi_i(t)} L_z \ket{\Psi_i(t)}$ and $\bra{\Psi_i(t)}S_z\ket{\Psi_i(t)}$ for $i=1,2$ stem from the population oscillation of $\braket{\Psi_i(t)}{\Psi_i(t)}$.  

The population oscillations' frequency can be estimated for a small perturbation $V=\delta V$ of an eigenstate $\Psi^{\pm}$
as the phase difference $S_1-S_2$ in Eq. \eqref{eq-phase-difference-ES} is no longer constant,
	\begin{eqnarray} \label{eq-Josefson-22}
	\dot{S}_1-\dot{S}_2 &=&  - 2 \, \delta  V \equiv  \omega.
	\end{eqnarray}
With $\Delta = \omega t$ we get for the lowest eigenstate ($q=0$) Josephson population oscillations as seen from Eq. \eqref{eq-population-time-ES} which becomes 
	\begin{eqnarray} 
	&& \dot{n}_1(\varphi) =         \kappa\sqrt{n_1n_2}  \sin[ \omega t]    ,\nonumber \\
	&& \dot{n}_2(\varphi) =   -   \kappa\sqrt{n_1n_2}   \sin [\omega t]  = - \dot{n}_1. \label{eq-population-time-JS} 
	\end{eqnarray}
These coupled differential equations can be solved exactly with $\mp$ for $n_{1,2}$ respectively, 
	\begin{eqnarray} \label{eq-Josefson-25}
	n_{1,2}(t)=  \Big( \sqrt{n_{1,2}(0)} \cos [ \kappa/ \omega \sin^2[  \omega t/2 ] ] \nonumber \\ 
	\mp \sqrt{n_{2,1}(0)} \sin[ k/ \omega \sin^2[  \omega t/2 ] ] \Big)^2 .
	\end{eqnarray}
Eq. \eqref{eq-Josefson-25} describes population oscillations with a frequency of $\omega =  2 \delta V$. The observed frequency in Fig \ref{pic-J-AC} is only slightly smaller then $\omega = 2 \delta V$ frequency.

\section{The spacial Josephson effect and vortex nucleation}

In solid state materials the phase difference  between the two macroscopic wave functions is defined as
the difference at the boundary surface of  bulk materials. Each phase is 
considered constant near the boundary surface and the phase inside the bulk material is assumed not to contribute.  
The situation in the ring offers a different physical scenario as the relevant phase difference is not necessarily the same in every point of the ring. 
%
%%%%%%%%%%%%%%%%%%%%%%%%%%%%%%%%%%%%%%%%%%%%%%%%%%%%%%%%%%%%%%%%%%%%%%%%%%%%%%%%%%%%
\begin{figure} [b,t,h]
	\includegraphics[width=0.45\textwidth]{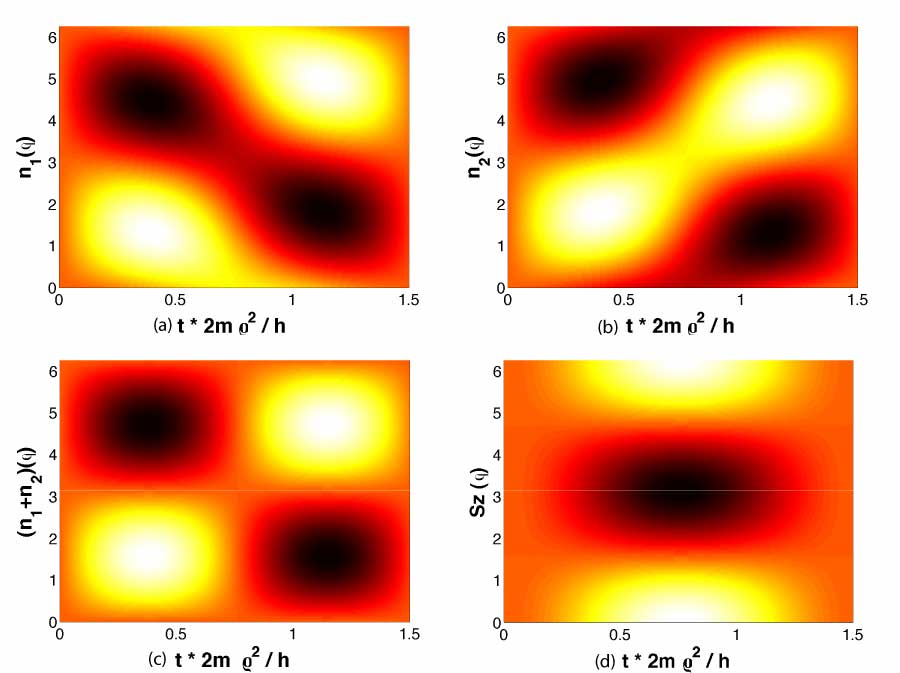}
 		\caption{ \label{pic-Josephson-local1-ring} Spacial Josephson oscillations: Single component density (a, b) and overall density (c) showing spacial oscillations for the initial state  $(1,-\I)$. The spin orbital angular momentum $S_{\mathrm{z}}(\varphi,t)$ oscillates locally, but not globally. (Compare with $S_{\mathrm{z}}(t)$ in Fig. \ref{pic-vortex-nucleation} (f)).
	}  
 \end{figure} 
%%%%%%%%%%%%%%%%%%%%%%%%%%%%%%%%%%%%%%%%%%%%%%%%%%%%%%%%%%%%%%%%%%%%%%%%%%%%%%%%%%%

\begin{figure} [b,t,h]
	\includegraphics[width=0.45\textwidth]{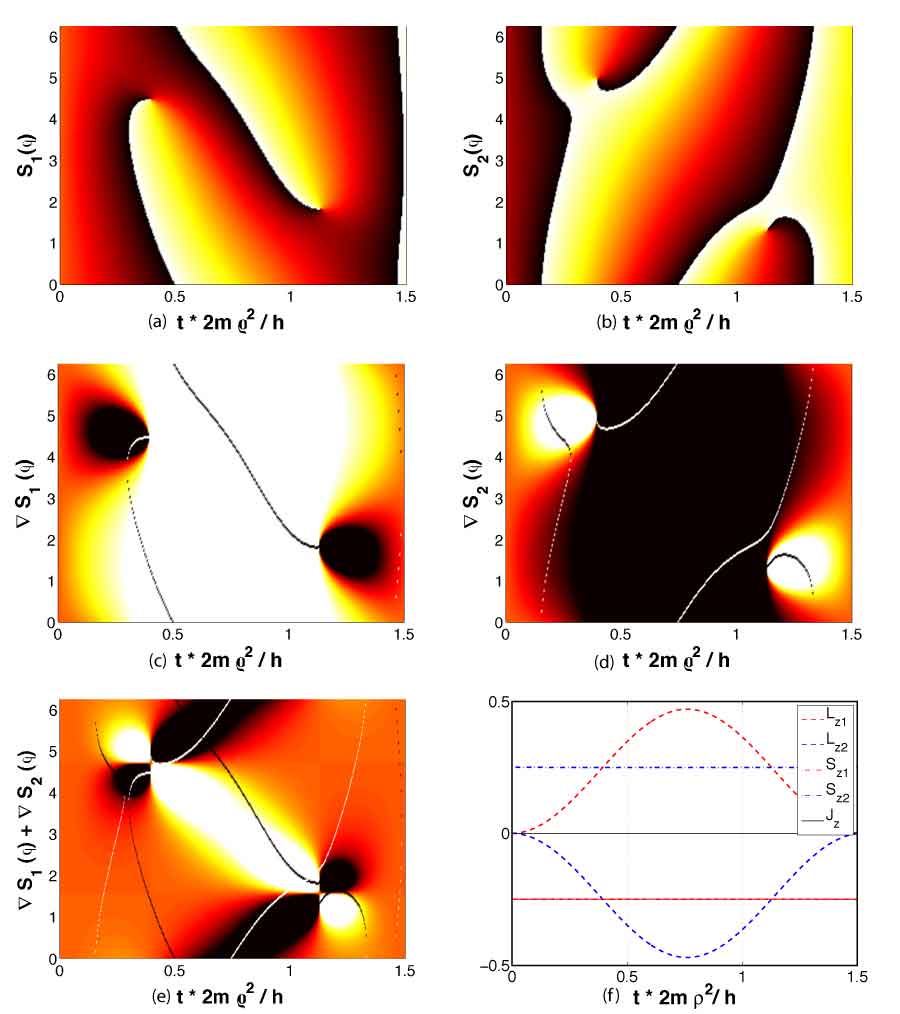}
		\caption{  \label{pic-vortex-nucleation}
	The phase and the phase gradient of each spinor component are depicted as function of time. The singularities in the phase-time diagram (a),(b) indicate vortex creation and annihilation. 
	The gradient of the phase is overall  positive (negative) sign once the  current is established  (white / black area in (c),(d)).
	The black and white lines are the $\pm 2\pi$ phase jumps. (f) The single component orbital angular momentum $L_{\mathrm{z}i}(t)$ oscillates while the total spin angular momentum $J_{\mathrm{z}}(t)$ is constant. } 
 \end{figure}

To illustrate this we consider an initial state of the form
	\begin{equation} \label{eq-ansatz-filled-r}
	\Psi= \frac{1}{ \sqrt {4 \pi} }
	\left( \begin{array}{c} 
	1\\
	-\I   
	\end{array} \right),		
	\end{equation}
which leads, for a non-interacting gas ($g=0$), to dynamics according to \eqref{eq-Josephson-short1}-\eqref{eq-Josephson-short4}
	\begin{eqnarray}
	&&  \dot{S}_1 =  - \kappa^2 +    \frac{ \kappa\sqrt{n_2} }{2 \sqrt{n_1} }  \cos \varphi, \label{eq-dynamic-spacial-vortex4} \\
	&& \dot{n}_1 (\varphi)=    -    \kappa\sqrt{n_1n_2} \sin \varphi   , \\
	&&  \dot{S}_2 =    - \kappa^2 -       \frac{ \kappa\sqrt{n_1} }{2  \sqrt{n_2} }\cos \varphi  , \\
	&& \dot{n}_2 (\varphi)=     -   \kappa\sqrt{n_1n_2}  \sin \varphi. \label{eq-dynamic-spacial-vortex1}
	\end{eqnarray}
These equations have now a nontrivial spacial dependence where the density changes with same sign for both components of the system. As we have $ \dot{n_i} \sim \sin \varphi$  we expect a spacial distribution with one maximum and one minimum on the ring. Moreover, the  time derivative of the velocity field is $\partial_\varphi \dot{S}_1 = -\partial_\varphi \dot{S}_2$ if $n_1 = n_2$. 
Fig. \ref{pic-Josephson-local1-ring} and \ref{pic-vortex-nucleation} show the evolution of \eqref{eq-ansatz-filled-r} in  time. Initially we see the dynamics described by Eqs. \eqref{eq-dynamic-spacial-vortex4}-\eqref{eq-dynamic-spacial-vortex1} with spacial sinusoidal   density oscillations $n_i \sim \sin \varphi$ and currents $ \partial_\varphi n_1 (\varphi)= - \partial_\varphi n_2 (\varphi) $.

For longer times, the ring is described by the coupled and nonlinear eqs. \eqref{eq-Josephson-long1}-\eqref{eq-Josephson-short4}  and we
 observe not only the spacial density distributions but also counter propagating currents.
For the chosen gauge potential the effective magnetic field stems from the non-Abelian part and is given by a constant field in the $z$-direction
	\begin{equation}
	\mathsf{B} = \nabla \times \mathsf{A} - \frac{\I}{2\hbar} (\mathsf{A} \times \mathsf{A}) 
	= \hbar \kappa^2 \left( \begin{array}{cc} 
	1 & 0 \\
	0 & -1  
	\end{array} \right) \mathbf{e}_{z},
	\end{equation}
where both spinor components start to rotate against each other. 

In the phase-time  diagram for $S(\varphi,t)$ in Fig. \ref{pic-vortex-nucleation}  the moment of establishing (and vanishing) a quantised circulation in each spin component can be found where the phase singularities appear. In the velocity field $v_i = \partial_\varphi S_i(\varphi, t)$ we can clearly observe regions with positive and negative velocities  $v_i$.

Persistent currents and vortex nucleation have been widely discussed in literature \cite{Raman_2001, Chiba_2008, Lundh_2003, Sinha_2001, Food_2001, scherer:110402,
woo:031604, Damski_2002, Murray_2007,song:033602} as the nucleation of rotation needs typically a dynamic instability. The two component spinor
gas offers the option, that the components rotate against each other. As the gas' topology 
has to be preserved, the rotation emerges in counter rotating pairs. The necessary break up
of the coherent wave function happens at the single component density minima,  $|\Psi_i(\varphi)|^2=0$, where it costs no energy to change the phase. 
Nevertheless, the condensate itself does not break into parts as the singularities in the phase-time diagram appear
simultaneously but spatially separated (see Fig. \ref{pic-vortex-nucleation} (e)) by approximately  $15^\circ$. In addition, the orbital angular momentum $L_{\mathrm{z}i}=\bra{\Psi_i (t)} - \I \partial_\varphi \ket{\Psi_i (t)}$ changes smoothly for  each component in time and has no discontinuities. It is still possible to define a preserved winding number $\bar{q}$ by counting the $\pm 2 \pi$ phase jumps with the proper sign for both spinor components. The spin orbital momentum, $S_{\mathrm{z}}(t)$, is conserved in time, but not locally as $|\Psi_1(\varphi, t)|^2-|\Psi_2(\varphi, t)|^2 \ne 0$, which is shown in fig. \ref{pic-Josephson-local1-ring} (d).

%%%%%%%%%%%%%%%%%%%%%%%%%%%%%%%%%%%%%%%%%%%%%%%%%%%%%%%%%%%%%%%%%%%%%%%%%%%%%%%%%%%
%%%%%%%%%%%%%%%%%%%%%%%%%%%%%%%%%%%%%%%%%%%%%%%%%%%%%%%%%%%%%%%%%%%%%%%

%%%%%% &&&&&&&&&&&&&& %%%%%%

\section{The role of interactions}

As shown in Section \ref{sec-Eigenstates} the eigenstates do not depend explicitly on the interaction strength $g$ and also the internal Josephson oscillations are not qualitatively affected by the interactions. In contrast the spacial oscillations and thereby especially the mass currents are much more fragile in the presence of interactions. This is indeed to be expected. The elementary excitations of a standard single component condensate generally also depend on the interaction, which for instance give the expression for the speed of sound in the gas.  An additional restriction which needs to be kept in mind for the spin-orbit coupled gas is the size of the chemical potential, which, in order to fulfill the adiabatic approximation so that the atoms remain in the dark state manifold, must be smaller than the total Rabi frequency.

It is first of all important to note that an arbitrary initial state can induce large amplitude oscillations and population transfer between the spin components. In the presence of interactions, and consequently nonlinearities, the dynamics is inevitably going to be complicated if the initial state is not similar to one of the eigenstates. This is indeed always the case, also for single component condensates, however, the spinor character of the ring studied here tends to enhance this effect due to the increased parameter space.  

In order to prepare a well defined coherent spinor condensate as initial state several criteria needs to be fulfilled. First of all, as mentioned above, the adiabaticity needs to be preserved. Secondly, the spin-orbit coupled gas is notoriously problematic due to fragmentation which would preclude the formation of a coherent condensate \cite{Stanescu2007}. To circumvent this problem one can prepare a well defined condensate in one of the three lowest $m=-1,0,+1$ states, and then transfer adiabatically the condensate to the dark state manifold. By doing so the coherence is expected to be preserved, and a well defined Gross-Pitaevskii equation can be used to describe the gas.

The resulting dynamics would however typically be well within the nonlinear regime, meaning that a Bogoliubov-de Gennes treatment, or in other words, a linearisation of the Gross Pitaevskii dynamics, is not appropriate. This is because the adiabatically transferred state is not necessarily going to be a steady state solution in the new spin-orbit coupled setting. Any attempt to identify eigenmodes and eigenfrequencies will therefore inevitably be strongly influenced by the collisional interactions. In order to analyse the elementary excitations using the Bogoliubov-de Gennes treatment  one should be close to the ground state of the system and study small deviations from it. But this is not necessarily straightforward because the fragmentation will prevent us from preparing, i.e., cooling, the condensate directly into a coherent ground state. Therefore, in most cases, there will be dynamics taking place, but some adiabatically prepared initial states will be better approximations of a coherent steady state than others. 
 
We have simulated the dynamics of the ring condensate using a single scattering length $g$. Strictly speaking scattering takes place between the bare electronic levels and not in the dressed state basis of the dark states $\Psi_i$. A more detailed analysis has to take different scattering lengths  between the atomic hyperfine levels of the tripod system into account and project these on the dark state basis. However, if the scattering lengths are close to each other, one can justify the usage of a single parameter $g$ \cite{Diss_Andreas}.  For $^{87}\mathrm{Rb}$ the scattering lengths are typically within a few percent   \cite{PhysRevLett.80.2097}.

\begin{figure} [b,t,h]
	\includegraphics[width=0.45\textwidth]{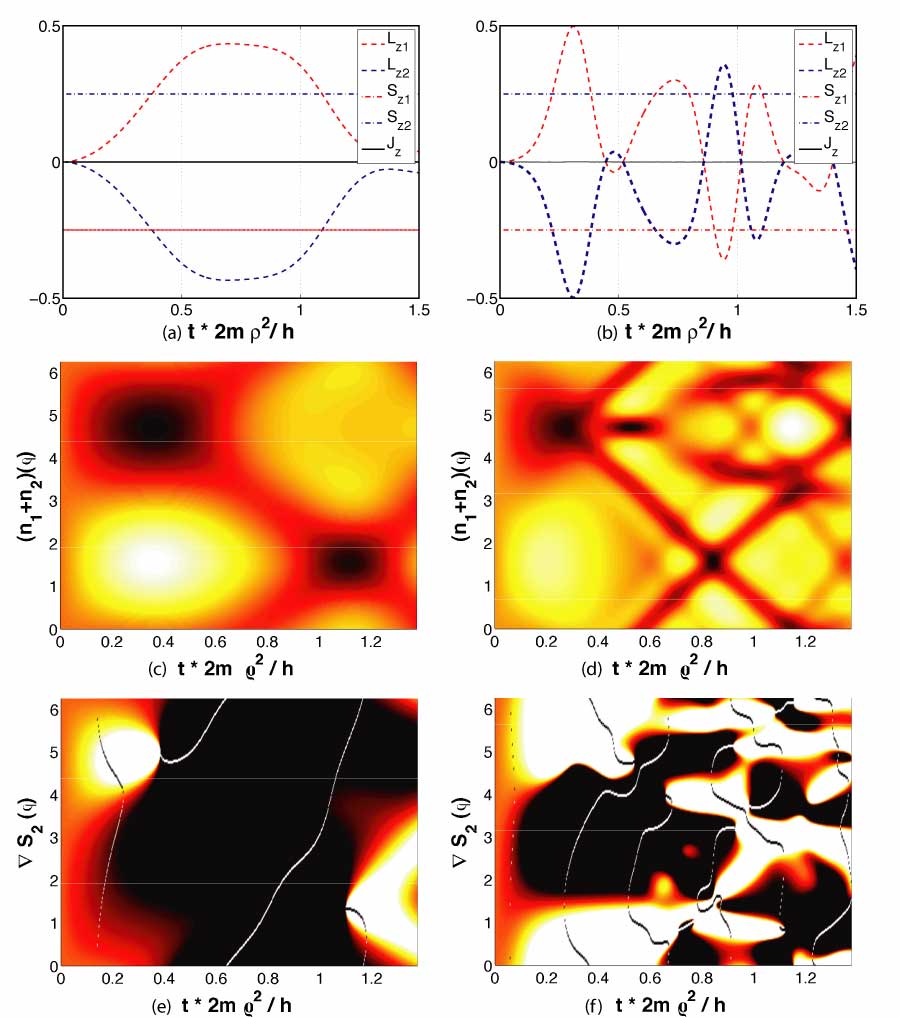}
	\caption{  \label{pic-interaction}
	Strong Interactions destroy the mass current. 
	(a,b) Total, orbital and spin angular momentum, (c,d) density, and (e,f) phase gradient with interactions for $g=2$ (left) and $g  =25 $ (right); Different than in the interactionless case $(g=0)$ a regime with persistent currents over the full ring is not established. The initial state is $1/\sqrt{4 \pi} \, (1,-\I)^{\mathrm{T}}$. 
} 
 \end{figure}

\begin{figure} [b,t,h]
	\includegraphics[width=0.45\textwidth]{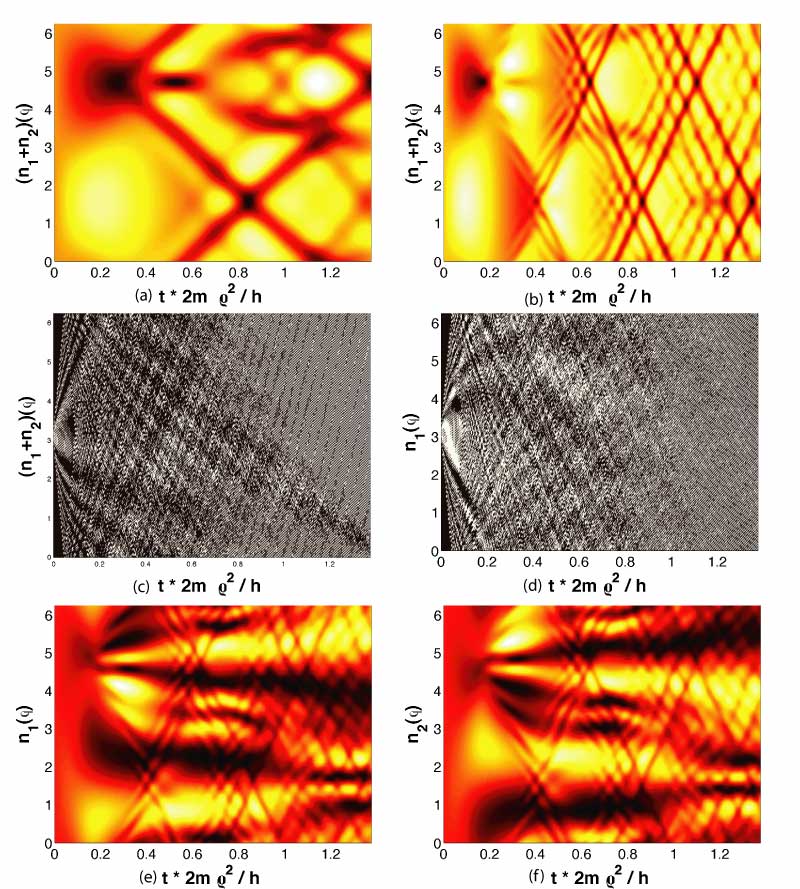}
	\caption{  \label {pic-grey-soliton}
	Density dynamics for $g=25$ (a,c) and $g=150 $ (b,d,e,f). Phonons can be seen to propagate at the speed of sound (middle), while the grey soliton like objects propagate considerably slower. 
In the strongly interacting regime the two condensate components start to separate (e, f). 
	}
\end{figure}

Our simulations show in Fig. \ref{pic-interaction} a significant influence on the mass current creation above an interaction strength for which $g >5$.                                                       
Instead of inducing counter propagating currents, the spacial Josephson oscillations lead to several pairs of soliton like objects which circulate around the ring \cite{PhysRevLett.86.2918}.  The non-Abelian magnetic field is not strong enough to induce stable currents any more (see Fig. \ref{pic-grey-soliton}).

%%%%%%%%%%%%%%%%%%%%%%%%%%%%%%%%%%%
%%%%%%%%%%%%%%%%%%%%%%%%%%%%%%%%%%%%%%%%%%%%%%%%

%%%%%%%%%%%%%%%%%%%%%%%%%%%%%%%%%%%
 %%%%%%%%%%%%%%%%%%%%%%%%%%%%%%%%%%%%%

\section{Summary}

The bosonic non-Abelian ring offers a rich toolbox for cold atom physics.
We investigated a homogeneously filled ring which can be fully described and understood in terms of the well known 
internal, nonlinear Josephson effect. Moreover, it was found that this mechanism can cause large  density oscillations and give rise to grey soliton-like structures.

The eigenstates were calculated exactly. They show mass currents  where detunings  
in the external trapping or deviations from the eigenstate population ratio leads to the AC-Josephson effect. 
Spacial Josephson oscillations appear  for the most simple 
configuration of an equally filled ring without any initial rotation. Already the non-interacting case shows non-trivial behaviour as  spacial density oscillations and oscillating mass currents appear. 
In the interacting case the coherent oscillations in the quasi-spin populations and the spin currents breaks down for sufficiently large interactions, which results in the appearance of soliton-like structures. 

Observing the spacial density oscillations would experimentally validate the
non-Abelian nature of the optically induced vector potential. 
In the dilute regime the nucleated rotation vanishes periodically but the found effect
might be used to initiate rotation in the system and switch off the laser
field in a coherent way to preserve the rotating states in their traps.
Furthermore we found a mechanism to nucleate persistent currents  without
any dynamical instability or atom-atom interaction necessary.

Finally it is worth noting that there is extensive discussions in the literature about  non-Abelian Josephson effects, which is
a field theoretical generalisation of the conventional Josephson effect and can be connected to
the Higgs mechanism \cite{esposito:241602, qi:185301}. In our paper,
we use a non-Abelian vector potential to couple the two wave function dynamically, but
the observed spacial Josephson effect is still a realisation of the Abelian Josephson effect. For the non-Abelian effect the setting is, e.g., a double well potential separating an Abelian and a non-Abelian region.

\acknowledgements

Helpful discussions with Julius Ruseckas are greatly acknowledged.
One of us (MM) acknowledge Peter Kr\"uger for his inspiring talk about  a BEC trapped on a 
torus at the Oxford Summer School 2008 on Cold Atoms in Optical Lattices, and the $\text{ECT}^*$ Doctoral Training Program 2009 in Trento where parts of this work were carried out.
This work was supported by the UK EPSRC. 

%\bibliography{articles-Rc6a}

\appendix
\section{Derivation of Josephson equation} \label{Appendix-Jos}

The set of nonlinear coupled equations  \eqref{eq-Josephson-long1} to \eqref{eq-hydro-2-density} is obtained from \eqref{eq-ansatz-hydro}
	\begin{equation}  \label{eq-app-1}
	\Psi_J^\dagger \I  \partial_t \Psi_J = \Psi_J^\dagger H_\varphi \Psi_J, 
	\end{equation}
with the hydrodynamic description of the condensate wave function, $(\Psi_1, \Psi_2)^{\rm T}=(n_1 \exponent{\I S_1}, - \I n_2 \exponent{\I S_2})^{\rm T} $.
For the left hand side of \eqref{eq-app-1} we use the relation 
	\begin{equation}\label{eq-josephson-lefthand1}
	\I  \partdiff{t} \sqrt{n_i} \exponent{\I S_i } = \I  \frac{ \exponent{\I S_i }}{2 \sqrt{n_i} } \partdiff{t} n_i  -  \sqrt{n_i} \exponent{\I S_i \varphi}  \partdiff{t} S_i 
	\end{equation}
After multiplication with $(\Psi_1^*, \Psi_2^*)$ from the left the time dependence reads
	\begin{equation} \label{eq-timedependence}
	\Psi_i^* \I  \partial_t \Psi_i = \I  \frac{1}{2} \dot{n}_i -  n_i \dot{S}_i.
	\end{equation}
On the right hand side of \eqref{eq-app-1} we have diagonal and off-diagonal parts.
The diagonal parts can be calculated with 
	\begin{eqnarray} \label{eq-diagonal1}
	&&-\frac{1}{2} \Psi_i^* \partial_\varphi^2 \Psi_i=   -\frac{\I}{2 } [ \partial_\varphi S_i \partial_\varphi n_i + n_i  (\partial_\varphi^2 S_i) ]  \nonumber \\
	&&-
	        \frac{1}{8  n_i} (\partial_\varphi n_i)^2 -
		  \frac{1}{4 } (\partial_\varphi^2 n_i)  +
		   \frac{n_i}{2 }   (\partial_\varphi S_i)^2 .
	\end{eqnarray}
For the off-diagonal terms, mixing the two components $\Psi_1, \Psi_2$, we use 
	\begin{eqnarray} \label{eq-of-diagonal}
	&&\sqrt{n_j} \exponent{- \I S_j} \partial_\varphi \sqrt{n_i} \exponent{ \I S_i} = \\
	 &&\exponent{ \I (S_i-S_j)} 1/2 \sqrt{ \frac{n_j}{n_i} } \partial_\varphi n_i + \I \sqrt{n_i n_j} \exponent{ \I (S_i-S_j)} \partial_\varphi S_i. \nonumber
	\end{eqnarray}
The first off diagonal matrix element has the form
	\begin{eqnarray} \label{eq-app-4}
	&&\Psi_1^* \exponent{-\I \varphi} \kappa [\frac{-\I }{2}  +  \partial_\varphi] \Psi_2  
	= { \kappa } \sqrt{n_1 n_2} \exponent{\I [S_2-\varphi-S_1]}  \nonumber \\
	&& \times \Big[ -\frac{1}{2} - \frac{ \I }{ 2 n_2 } \partial_\varphi n_2 + \partial_\varphi S_2 \Big], 
	\end{eqnarray}
while the second off diagonal element reads
	\begin{eqnarray} \label{eq-result-second-offdiag}
	&&\Psi_2^* \exponent{\I \varphi} \kappa  [-\frac{\I }{2}  -  \partial_\varphi] \Psi_1 =  \kappa \sqrt{n_1 n_2}    \exponent{-\I [S_2-\varphi - S_1]} \nonumber  \\ 
	&& \times  \Big[ +  \frac{1 }{2 } - \frac{\I  }{ 2 n_1 }  \partial_\varphi n_1 +  \partial_\varphi S_1 \Big] .
	\end{eqnarray}
Collecting the real and imaginary parts of \eqref{eq-timedependence}, \eqref{eq-diagonal1} , \eqref{eq-app-4} and \eqref{eq-result-second-offdiag} gives using $\exponent{\I \Phi} = \I \sin\Phi + \cos \Phi$ the  coupled nonlinear equations for 
 the phases $S_i (\varphi, t)$ and the local densities $n_i (\varphi, t)$.

\end{document}